\documentclass[showpacs,preprint]{revtex4-1}
\usepackage{amssymb}
\usepackage{amsmath}
\usepackage{graphicx}

\begin{document}
\title{Diffuse approximation to the kinetic theory in a Fermi system}
\author{V.M. Kolomietz and S.V. Lukyanov}
\affiliation{Institute for Nuclear Research, 03680 Kyiv, Ukraine}
\pacs{21.60.-n, 21.60.Ev, 24.30.Cz}

\begin{abstract}
We suggest the diffuse approach to the relaxation processes within the
kinetic theory for the Wigner distribution function. The diffusion and drift
coefficients are evaluated taking into consideration the interparticle
collisions on the distorted Fermi surface. Using the finite range
interaction, we show that the momentum dependence of the diffuse coefficient
$D_{p}(p)$ has a maximum at Fermi momentum $p=p_{F}$ whereas the drift
coefficient $K_{p}(p)$ is negative and reaches a minimum at $p\approx p_{F}$. 
For a cold Fermi system the diffusion coefficient takes the non-zero value
which is caused by the relaxation on the distorted Fermi-surface at
temperature $T=0$. The numerical solution of the diffusion equation was
performed for the particle-hole excitation in a nucleus with $A=16$. The
evaluated relaxation time $\tau_{r}\approx 8.3\cdot 10^{-23}\mathrm{s}$ is
close to the corresponding result in a nuclear Fermi-liquid obtained within
the kinetic theory.
\end{abstract}

\keywords{Kinetic theory; Fermi liquid; diffusion approximation; relaxation time; 
particle-hole distortion.}

\maketitle

\section{Introduction}

The relaxation processes in many-body systems can be effectively studied
within the kinetic theory, see Ref. \cite{kosh04}\ and references therein.
The kinetic approaches operate with the kinetic equation, which is written
for the distribution function in phase space. The advantage of the kinetic
approaches is that the kinetic equation can be easily generalized to the
case of finite temperatures. Certain difficulties arise when one tries to
describe the relaxation and damping effects involving the collision integral
in $9$-dimension space \cite{bert78,kopl95,mako95,kiko96,kopl96,diko99}.

To reduce the kinetic equation for the Wigner distribution function, we will
follow the diffuse approach \cite{lipi81} considering the relaxation on the
distorted Fermi surface. The presence of the\textrm{\ }Fermi-surface
distortion effects gives rise to some important consequences. Because of the
Fermi-surface distortion, the scattering of particles leads to the
relaxation and the damping. The purpose of present paper is to study the
diffusion and drift terms in a Fermi-system which can be applied to the
description of the damping of collective and particle-hole excitations as
well as the large amplitude dynamics. In Sec. 2 we consider the kinetic
equation for the Wigner distribution function and reduce it applying the
diffusion approximation. In Sec. 3 we establish the diffusion and drift
coefficients taking into consideration the Fermi-surface distortion effects.
The discussion of numerical results is presented in Sec. 4. Our conclusions
are given in Sec. 5.

\section{\textbf{Collision integral within diffuse approximation}}

We will restrict ourselves to the Born collision approximation in the
kinetic equation for Wigner distribution function $f(\mathbf{r},\mathbf{p}%
;t)\equiv f$ \cite{kaba62,risc80}. Introducing the collision integral 
$\mathrm{St}\{f\}$, we will write the kinetic equation as
\begin{equation}
{\frac{\partial f(\mathbf{r},\mathbf{p}_{1};t)}{\partial t}}+\hat{L}f(
\mathbf{r},\mathbf{p}_{1};t)=\mathrm{St}\{f\},  \label{lv}
\end{equation}%
where $\hat{L}$ \ is the driving operator. In the lowest order in $\hbar$
the driving operator $\hat{L}$ is given by \cite{kosh04}
\begin{equation}
\hat{L}={\frac{1}{m}}\,\mathbf{p}\cdot \mathbf{\nabla }_{\mathbf{r}}-(%
\mathbf{\nabla }_{\mathbf{r}}U)\cdot \mathbf{\nabla }_{\mathbf{p}}.
\label{2}
\end{equation}%
Here, the single-particle potential $U$ includes, in general, the
self-consistent and external fields. The collision integral $\mathrm{St}%
\{f\} $ in Eq. (\ref{lv}) can be written in the following form \cite{abkh59}
\begin{eqnarray}
\mathrm{St}\{f\} &=&\int \frac{g^{2}d\mathbf{p}_{2}d\mathbf{p}_{3}d\mathbf{p}%
_{4}}{(2\pi \hbar )^{6}}\ \mathcal{W}(\{\mathbf{p}_{j}\})  \nonumber \\
&&\times \left[ \tilde{f}(\mathbf{p}_{1})\tilde{f}(\mathbf{p}_{2})f(\mathbf{p%
}_{3})f(\mathbf{p}_{4})-f(\mathbf{p}_{1})f(\mathbf{p}_{2})\tilde{f}(\mathbf{p%
}_{3})\tilde{f}(\mathbf{p}_{4})\right] \delta (\Delta \varepsilon )\delta
(\Delta \mathbf{p})  \nonumber \\
&=&\int \frac{gd\mathbf{p}_{3}}{(2\pi \hbar )^{3}}\ \left[ W_{3\rightarrow
1}(\mathbf{p}_{1},\mathbf{p}_{3})\tilde{f}(\mathbf{p}_{1})f(\mathbf{p}%
_{3})-W_{1\rightarrow 3}(\mathbf{p}_{1},\mathbf{p}_{3})f(\mathbf{p}_{1})%
\tilde{f}(\mathbf{p}_{3})\right] ,  \label{st1}
\end{eqnarray}%
where $g$ is the spin-isospin degeneracy factor, $\ f(\mathbf{p})\equiv f(%
\mathbf{r},\mathbf{p},t)$, $\tilde{f}(\mathbf{p})=1-f(\mathbf{p})$, $\Delta
\varepsilon =\varepsilon _{1}+\varepsilon _{2}-\varepsilon _{3}-\varepsilon
_{4},$ $\Delta \mathbf{p}=\mathbf{p}_{1}+\mathbf{p}_{2}-\mathbf{p}_{3}-%
\mathbf{p}_{4}$ and $\mathcal{W}(\{\mathbf{p}_{j}\})$ is the probability of
two-body collisions. The gain and loss terms $W_{3\leftrightarrows 1}(%
\mathbf{p}_{1},\mathbf{p}_{3})$ in Eq. (\ref{st1}) are given by
\begin{eqnarray}
W_{3\rightarrow 1}(\mathbf{p}_{1},\mathbf{p}_{3}) &\equiv &W(\mathbf{p}_{1},%
\mathbf{p}_{3})=\int \frac{gd\mathbf{p}_{2}d\mathbf{p}_{4}}{(2\pi \hbar )^{3}%
}\ \mathcal{W}(\{\mathbf{p}_{j}\})\tilde{f}(\mathbf{p}_{2})f(\mathbf{p}%
_{4})\delta (\Delta \varepsilon )\delta (\Delta \mathbf{p}),  \label{gains}
\\
W_{1\rightarrow 3}(\mathbf{p}_{1},\mathbf{p}_{3}) &\equiv &\widetilde{W}(%
\mathbf{p}_{1},\mathbf{p}_{3})=\int \frac{gd\mathbf{p}_{2}d\mathbf{p}_{4}}{%
(2\pi \hbar )^{3}}\ \mathcal{W}(\{\mathbf{p}_{j}\})f(\mathbf{p}_{2})\tilde{f}%
(\mathbf{p}_{4})\delta (\Delta \varepsilon )\delta (\Delta \mathbf{p}).
\label{losses}
\end{eqnarray}

The transition probability $W_{p\leftrightarrows q}(\mathbf{p},\mathbf{q})$
in Eq. (\ref{st1}) contains the square of the corresponding amplitude of
scattering for the direct, $1\rightarrow 3$, and the reverse, $3\rightarrow
1 $, transitions. The probability $W_{p\leftrightarrows q}(\mathbf{p},%
\mathbf{q})$ includes also the distribution functions of the scattered
particle in the initial and final states. We will assume that the main
contribution to the scattering amplitude is given by the transitions which
correspond to a small momentum transfer: $|\mathbf{p}_{1}-\mathbf{p}_{3}|\
\ll p_{F}$, where $p_{F}$ is the Fermi momentum, see also Ref. \cite{abkh59}%
. Introducing the new variables
\[
\mathbf{s}=\mathbf{p}_{3}-\mathbf{p}_{1}\quad \text{and}\quad \mathbf{P}=%
\frac{1}{2}(\mathbf{p}_{1}+\mathbf{p}_{3})=\mathbf{p}_{1}+\frac{\mathbf{s}}{2%
},
\]%
we will apply the following expansions over small $\mathbf{s}$ :
\begin{equation}
f(\mathbf{p}_{3})=f(\mathbf{p}_{1}+\mathbf{s})\approx f(\mathbf{p}%
_{1})+s_{\nu }\nabla _{p_{1},\nu }\ f(\mathbf{p_{1}})+\frac{1}{2}\ s_{\nu
}s_{\mu }\nabla _{p_{1},\nu }\nabla _{p_{1},\mu }\ f(\mathbf{p_{1}}),
\label{f3}
\end{equation}%
\begin{eqnarray}
W(\mathbf{p}_{1},\mathbf{p}_{3}) &=&W(\mathbf{P},\mathbf{s})  \nonumber \\
&\approx &W(\mathbf{p}_{1},\mathbf{s})+\frac{1}{2}\ s_{\nu }\nabla
_{p_{1},\nu }W(\mathbf{p}_{1},\mathbf{s})+\frac{1}{8}\ s_{\nu }s_{\mu
}\nabla _{p_{1},\nu }\nabla _{p_{1},\mu }W(\mathbf{p}_{1},\mathbf{s}),
\label{w13}
\end{eqnarray}%
and
\begin{eqnarray}
\widetilde{W}(\mathbf{p}_{1},\mathbf{p}_{3}) &=&\widetilde{W}(\mathbf{P},%
\mathbf{s})  \nonumber \\
&\approx &\widetilde{W}(\mathbf{p}_{1},\mathbf{s})+\frac{1}{2}\ s_{\nu
}\nabla _{p_{1},\nu }\widetilde{W}(\mathbf{p}_{1},\mathbf{s})+\frac{1}{8}\
s_{\nu }s_{\mu }\nabla _{p_{1},\nu }\nabla _{p_{1},\mu }\widetilde{W}(%
\mathbf{p}_{1},\mathbf{s}).  \label{w31}
\end{eqnarray}%
Note that the spin-averaged probability of two-body collisions $\mathcal{W}%
(\{\mathbf{p}_{j}\})$ in Eq. (\ref{st1}) can be expressed in term of
in-medium scattering cross section $d\sigma /d\Omega $\ as
\begin{equation}
\mathcal{W}(\{\mathbf{p}_{j}\})=\frac{2(2\pi \hbar )^{3}}{m^{2}}\frac{%
d\sigma }{d\Omega }(\{\mathbf{p}_{j}\}).  \label{wdsdo}
\end{equation}%
In the case of elastic collisions, the scattering cross section $d\sigma /d\Omega$ 
depends on the modulus square of the momentum transfer $\mathbf{s}$ only.

Using the expansions Eqs. (\ref{w13}) and (\ref{w31}), one can reduce the
kinetic equation (\ref{lv}) to the diffusion equation in the following form
(see Eq. (\ref{stf6}) of\textrm{\ Appendix A})
\begin{equation}
{\frac{\partial f}{\partial t}}+\hat{L}f=-\mathbf{\nabla }_{p}\left[ \mathbf{%
K}_{p}(\mathbf{p})f(\mathbf{p})\tilde{f}(\mathbf{p})+f^{2}(\mathbf{p})%
\mathbf{\nabla }_{p}D_{p}(\mathbf{p})\right] +\mathbf{\nabla }_{p}^{2}\,%
\left[ f(\mathbf{p})\,D_{p}(\mathbf{p})\right] ,  \label{4}
\end{equation}
where $D_{p}(\mathbf{p})$ and $K_{p}(\mathbf{p})$ represent the diffusion
and drift terms, respectively. Both kinetic coefficients $D_{p}(\mathbf{p})$
and $K_{p}(\mathbf{p})$ are derived by the following relations, see 
\textrm{Appendix A},
\begin{equation}
B_{\nu \mu }(\mathbf{p})=D_{p}(\mathbf{p})\delta _{\nu \mu },\quad D_{p}(%
\mathbf{p})=\frac{1}{6}\int \frac{gd\mathbf{s}}{(2\pi \hbar )^{3}}\ s^{2}\ W(%
\mathbf{p},\mathbf{s})  \label{dpdef1}
\end{equation}%
and
\begin{equation}
K_{p}(\mathbf{p})\nabla _{p,\nu }\varepsilon _{p}=\nabla _{p,\mu }B_{\nu \mu
}(\mathbf{p})-A_{\nu }(\mathbf{p})\equiv \nabla _{p,\nu }D_{p}(\mathbf{p}%
)-A_{\nu }(\mathbf{p}).  \label{kpdef1}
\end{equation}

\section{Kinetic coefficients}

The obtained expressions (\ref{dpdef1}) and (\ref{kpdef1}) for the kinetic
coefficients imply a smallness of the momentum transfer $\mathbf{s}$ because
of the expansions Eqs. (\ref{f3}), (\ref{w13}) and (\ref{w31}). To provide a
small momentum transfer $\mathbf{s}$ in Eqs. (\ref{dpdef1}) and (\ref{kpdef1}%
) we will use the finite-radius inter-particle interaction with the
following Gaussian form-factor $v(r)=v_{0}\exp (-r^{2}/2r_{0}^{2})$ which is
appropriate for calculations of the in-medium cross-section within the
transport approaches\textrm{\ }\cite{shda03,coly11}\textrm{.} The
differential cross section $d\sigma /d\Omega $ in the first Born
approximation is then given by \cite{Dav.b.65}
\begin{equation}
\frac{d\sigma (\{\mathbf{p}_{j}\})}{d\Omega }=\frac{\pi
m^{2}r_{0}^{6}v_{0}^{2}}{2\hbar ^{4}}\exp \left( -4\mathbf{s}%
^{2}r_{0}^{2}/\hbar ^{2}\right) ,  \label{pot}
\end{equation}%
were $r_{0}$ and $v_{0}$ are the free parameters.

\subsection{Diffusion term}

Using Eqs. (\ref{gain1}) and (\ref{pot}), we will rewrite the diffuse term $%
D_{p}(\mathbf{p}_{1})$ of Eq. (\ref{dpdef1}) as
\begin{eqnarray}
D_{p}(\mathbf{p}_{1}) &\approx &\frac{g^{2}r_{0}^{6}v_{0}^{2}}{48\pi
^{2}\hbar ^{7}}\int d\mathbf{p}_{2}d\mathbf{p}_{4}d\mathbf{s}\ s^{2}\exp
\left( -4\mathbf{s}^{2}r_{0}^{2}/\hbar ^{2}\right) \tilde{f}(\mathbf{p}%
_{2})f(\mathbf{p}_{4})\ \delta \left( \mathbf{p}_{2}-\mathbf{p}_{4}-\mathbf{s%
}\right)  \nonumber \\
&\times &\delta \left( \varepsilon _{2}-\varepsilon _{4}-\frac{\mathbf{p}_{1}%
\mathbf{s}}{m}\right) .  \label{dp}
\end{eqnarray}%
Integrating in Eq. (\ref{dp}) over $\mathbf{s}$, we obtain (see \textrm{%
Appendix B})
\begin{eqnarray}
D_{p}(p) &\approx &\frac{g^{2}mr_{0}^{6}v_{0}^{2}}{24\pi ^{2}\hbar ^{7}}%
\int_{0}^{\infty }dk\ k^{5}\int_{-1}^{1}dx\int_{-1}^{1}dy\ \exp \left[
-8k^{2}r_{0}^{2}(1-xy)/\hbar ^{2}\right]  \nonumber \\
&\times &\left\{ (1-xy)\ j\left( 8k^{2}r_{0}^{2}\sqrt{1-x^{2}}\sqrt{1-y^{2}}%
/\hbar ^{2}\right) -\sqrt{1-x^{2}}\sqrt{1-y^{2}}\right.  \nonumber \\
&\times &\left. \widetilde{j}\left( 8k^{2}r_{0}^{2}\sqrt{1-x^{2}}\sqrt{%
1-y^{2}}/\hbar ^{2}\right) \right\} \tilde{f}\left( \sqrt{k^{2}+p^{2}+2kpx}%
\right)  \nonumber \\
&\times &f\left( \sqrt{k^{2}+p^{2}+2kpy}\right) ,  \label{dp8x}
\end{eqnarray}%
where $x=\cos \theta _{q}$ and $y=\cos \theta _{k}$.

We will apply our consideration to the nuclear Fermi-liquid. For the
numerical calculations we will adopt the following parameters $r_{0}=0.8$
\textrm{fm} and $v_{0}=-33$ \textrm{MeV}, which provide a reasonable value
for the in-medium nucleon-nucleon cross section $\sigma _{\mathrm{tot}%
}\simeq 20$ \textrm{mb}. We will also use the Fermi distribution function
\begin{equation}
f(p)=\left( 1+\exp \frac{p^{2}/2m-\lambda (T)}{T}\right) ^{-1},  \label{fd}
\end{equation}%
where $T$ is the temperature, $\lambda (T)\approx \varepsilon _{F}\left[
1-(\pi ^{2}/12)(T/\varepsilon _{F})^{2}\right] $\ is the chemical potential\
and $\varepsilon _{F}=37$ \textrm{MeV} is the Fermi energy.

\begin{figure}
\begin{center}
\includegraphics[scale=0.4,clip]{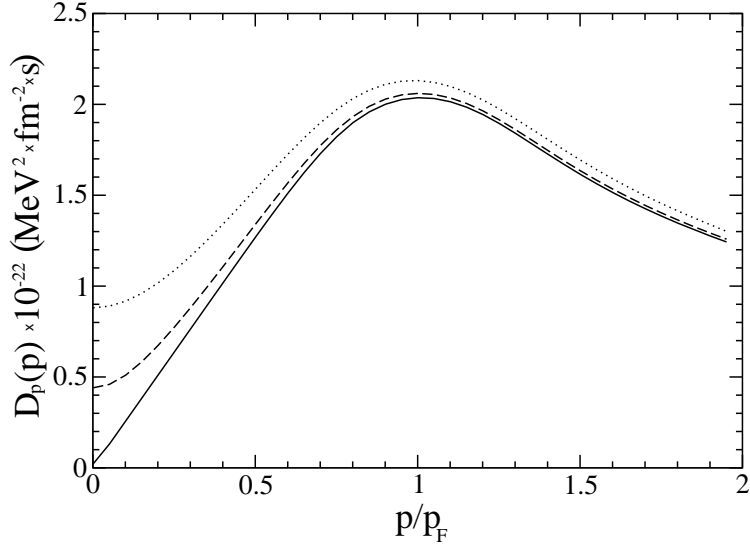}
\end{center}
\caption{Dependence of the diffusion coefficient $D_{p}(p)$ on momentum $p$ in
units of $p_{F}$, for temperatures $T=0.1$ \textrm{MeV} (solid line), $2\
\mathrm{MeV}$ (dashed line) and $4$ $\mathrm{MeV}$ (dotted line).}
\label{fig1}
\end{figure}

The results of calculations of the diffusion coefficient $D_{p}(p)$
accordingly to Eq. (\ref{dp8x}) with the Fermi distribution (\ref{fd}) are
presented in \textrm{Fig. 1}. The calculations were performed for the
different values of temperature $T$. As seen from \textrm{Fig. 1}, the
momentum dependence of $D_{p}(p)$ has the clearly observed maximum at the
Fermi momentum $p_{F}$ for different temperatures. With an increase of
temperature the diffusion coefficient increases as $D_{p}(p=p_{F},T)\sim
T^{2}$. The temperature dependence of the diffusion coefficient $D_{p}(p=p_{F},T)$ 
is shown in \textrm{Fig. 2.} As seen from \textrm{Fig. 2}, for a cold 
Fermi system the diffusion coefficient takes the
non-zero value $D(p_{F},T=0)=2.1\cdot 10^{-22}$ $\mathrm{MeV}^{2}\mathrm{
\cdot fm}^{-2}\mathrm{\cdot s}$ which is caused by the relaxation on the
distorted Fermi-surface in Eq. (\ref{lv}) which exists at $T=0$ also. For
high temperature regime where the Fermi statistic comes to Maxwell one, the
diffusion coefficient $D(p_{F},T)$ behaves as a linear function of $T$ in
agreement with the Einstein's fluctuation-dissipation theorem.
\begin{figure}
\begin{center}
\hspace{1cm}\includegraphics[scale=0.4,clip]{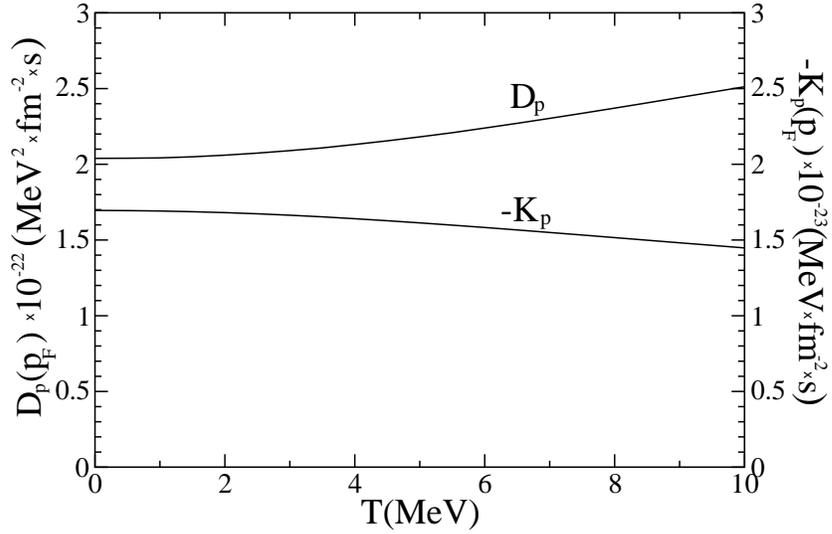}
\end{center}
\caption{Dependence of the diffusion, $D_{p}$, and drift, $K_{p}$,
coefficients taken at $p=p_{F}$ on the temperature $T$.}
\label{fig2}
\end{figure}

\subsection{Drift term}

Using Eq. (\ref{kpdef1}) and $\varepsilon _{p}=p^{2}/2m$, one can write the
drift coefficient $K_{p}(\mathbf{p})$ as
\begin{equation}
K_{p}(\mathbf{p})\frac{p_{\nu }}{m}=\nabla _{p,\nu }D_{p}(\mathbf{p})-%
\widehat{p}_{\nu }A(\mathbf{p}),  \label{K1}
\end{equation}%
where
\begin{equation}
A(\mathbf{p})=\frac{1}{p}\int \frac{gd\mathbf{s}}{(2\pi \hbar )^{3}}\ p_{\nu
}s_{\nu }\ W(\mathbf{p},\mathbf{s}).  \label{A1}
\end{equation}%
For a spherically symmetric distribution $f(p)$ the drift coefficient $K_{p}(%
\mathbf{p})$ in Eq. (\ref{K1}) is reduced to the following form (see \textrm{%
Appendix B}) \
\begin{equation}
K_{p}(p)=\frac{m}{p}\left( \frac{\partial D_{p}(p)}{\partial p}-A(p)\right) ,
\label{kpp}
\end{equation}%
where the first moment function $A(p)$ is given by Eq. (\ref{dk8x})
of \textrm{Appendix B}.

The presence of maximum of $D_{p}(p)$ at the Fermi momentum denotes that $%
\left. \partial D(p)/\partial p\right\vert _{p=p_{F}}=0$ and in agreement
with Eq. (\ref{kpp}) the drift coefficient $K_{p}(p_{F})$ is reduced as
\begin{equation}
K_{p}(p_{F})=-\frac{m}{p_{F}}A(p_{F}).  \label{kppF}
\end{equation}%
In \textrm{Fig. 3} we show the dependence of the drift coefficient $K_{p}(p)$
on the momentum $p$ in units of Fermi momentum $p_{F}$. One can see that $%
K_{p}(p)$ has a minimum localized at $p<$ $p_{F}$ which slightly depends on
the temperature. Note that for $p>p_{F}$ the temperature dependence of $%
K_{p}(p)$ is negligible.

\begin{figure}
\begin{center}
\includegraphics[scale=0.4,clip]{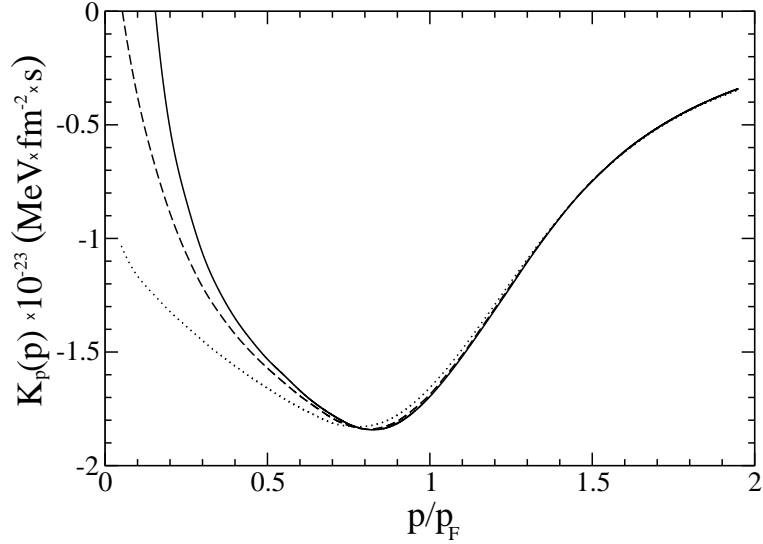}
\end{center}
\caption{The same as in \textrm{Fig. 1} but for the drift term $K_{p}(p)$.}
\label{fig3}
\end{figure}

The numerical estimates of the diffuse coefficient $D_{p}(p)$ and the drift term
$K_{p}(p)$ near the Fermi momentum give $D_{p}(p_{F})\approx 10^{-22}$ $%
\mathrm{MeV}^{2}\mathrm{\cdot fm}^{-2}\cdot \mathrm{s}$ and $%
K_{p}(p_{F})\approx -10^{-23}$ $\mathrm{MeV\cdot fm}^{-2}\cdot \mathrm{s}$.
Both obtained values of $D_{p}(p_{F})$ and $K_{p}(p_{F})$ agree with the
phenomenological ones used earlier in Ref. \cite{wols82}.

\section{Numerical results}

\bigskip

Evaluating the transport coefficients $D_{p}(p)$ and $K_{p}(p)$ caused by
the interparticle collisions on the distorted Fermi surface, we will
restrict ourselves to a Fermi system which is homogeneous in $\mathbf{r}$
-space and omit the driving operator $\hat{L}$ in equation (\ref{4}).
Using the diffusion, $D_{p}(p)$, and drift, $K_{p}(p)$,
coefficients from Eqs. (\ref{dp8x}) and (\ref{kpp}), and solving the
diffusion equation (\ref{4}) with $\hat{L}=0$ one can
evaluate the time evolution of the\ Wigner distribution function $f(p,t)$.
We will also assume a spherical Fermi surface of radius $p_{F}$ which is
derived by the condition for the particle number $A$ within a fixed volume $%
\mathcal{V}$
\[
\int_{0}^{p_{F}}\frac{4\pi g\mathcal{V}}{(2\pi \hbar )^{3}}\ p^{2}dp=A.
\]%
The\ diffusion equation (\ref{4}) must be augmented\ by the initial
condition for $f(p,t)$.\ \ We will consider the time evolution of the
initial particle-hole ($1p1h$) excitation which is derived at $t=0$ as, see
also Ref. \cite{wols82},
\begin{eqnarray}
f_{\mathrm{in}}(p) &=&\left[ 1-\theta (p-p_{1}^{\prime })+\theta
(p-p_{2}^{\prime })\right] \left[ 1-\theta (p-p_{F})\right]  \nonumber \\
&&+\left[ 1-\theta (p-p_{2})\right] \theta (p-p_{1})\theta (p-p_{F}).
\label{fin}
\end{eqnarray}%
The distribution $f_{\mathrm{in}}(p)$ of Eq. (\ref{fin}) means the particle
located at $p_{1}<p<p_{2}$ and the hole excitation at $p_{1}^{\prime
}<p<p_{2}^{\prime }$ for fixed $p_{1}>p_{F}$ and $p_{2}^{\prime }<p_{F}$,
respectively. The intervals $\Delta p^{\prime }=p_{2}^{\prime
}-p_{1}^{\prime }$ and $\Delta p=p_{2}-p_{1}$ are derived from the
conditions
\begin{eqnarray}
\int_{0}^{p_{F}}\frac{4\pi g\mathcal{V}dp}{(2\pi \hbar )^{3}}p^{2}f_{\mathrm{%
in}}(p,t=0) &=&A-1,  \nonumber \\
\int_{p_{F}}^{\infty }\frac{4\pi g\mathcal{V}dp}{(2\pi \hbar )^{3}}p^{2}f_{%
\mathrm{in}}(p,t=0) &=&1.  \label{fin2}
\end{eqnarray}

In \textrm{Fig. 4} we have plotted the time evolution of the Wigner
distribution function $f(p,t)$ for the initial particle-hole excitation in
the Fermi-system with $A=16$. We used the initial distribution $f_{\mathrm{in%
}}(p,t=0)$ given by Eq. (\ref{fin}) and assumed the initial excitation
energy $E_{\mathrm{ex}}=30$ $\mathrm{MeV}$.

\begin{figure}
\begin{center}
\includegraphics[scale=1.2,clip]{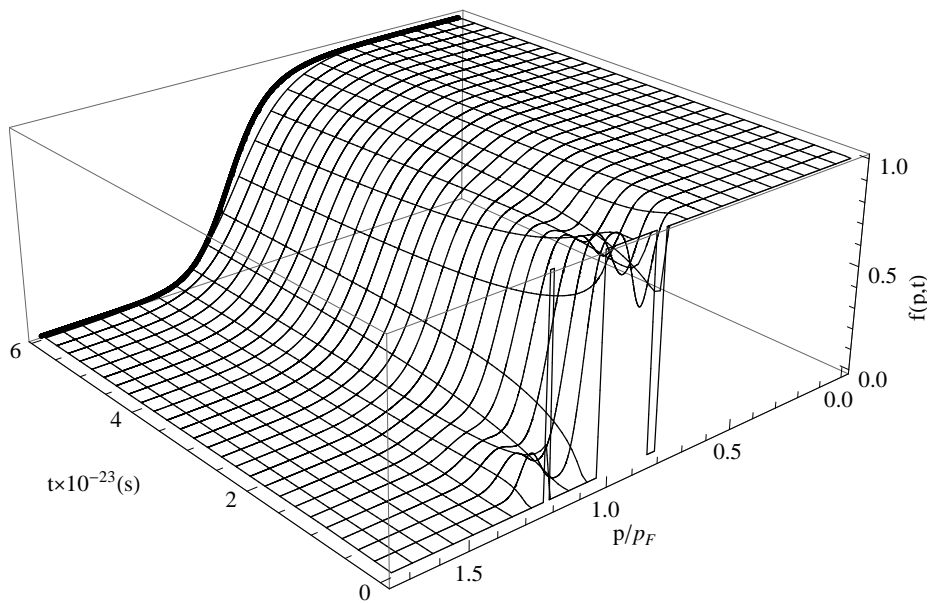}
\end{center}
\caption{Time evolution of initial distribution function (\protect\ref{fin})
in momentum space (in units of the Fermi momentum $p_{F}$) with $%
p_{1}^{\prime }/p_{F}\simeq 0.71$, $p_{2}^{\prime }/p_{F}\simeq 0.75$, $%
p_{1}/p_{F}\simeq 1.12$, $p_{2}/p_{F}\simeq 1.13$, which corresponds to the
initial excitation energy $E_{\mathrm{ex}}=30$ MeV. The solid line is the
equilibrium distribution $f_{\mathrm{eq}}(p)$.}
\label{fig4}
\end{figure}
One can see from \textrm{Fig. 4} that the momentum distribution $f(p,t)$
evolves to the Fermi equilibrium limit $f_{\mathrm{eq}}(p)$ of Eq. (\ref{fd}) 
with temperature $T=\sqrt{E_{\mathrm{ex}}/a}$, where $a=(\pi
^{2}/6)g(\epsilon _{F})$\ is the level density parameter \cite{bomo1} and $%
g(\epsilon _{F})$\ is the single-particle level density at the Fermi energy%
\textbf{.} The corresponding relaxation time $\tau _{r}$ for the diffusion
process in \textrm{Fig. 4} can be obtained considering the time evolution of
the deviation $\delta f(p,t)=f(p,t)-f_{\mathrm{eq}}(p)$ of the distribution
function $f(p,t)$ from its equilibrium limit $f_{\mathrm{eq}}(p)$. We
introduce the mean square deviation
\begin{equation}
\Delta (t)=\int d\mathbf{p\ [}\delta f(p,t)\mathbf{]}^{2}-\left[ \int d%
\mathbf{p\ }\delta f(p\mathbf{,}t\mathbf{)}\right] ^{2}=\int d\mathbf{p\ [}%
\delta f(p,t)\mathbf{]}^{2}.  \label{deltat}
\end{equation}%
The time dependence of $\Delta (t)$ for the distribution function $f(p,t)$
from \textrm{Fig. 4} is plotted in \textrm{Fig. 5}. The function $\Delta (t)$
can be fitted by the exponential dependence $\Delta (t)\sim $ $\exp (-t/\tau
_{r})$ which is shown in \textrm{Fig. 5} as the dashed line.
\begin{figure}
\begin{center}
\includegraphics[scale=0.4,clip]{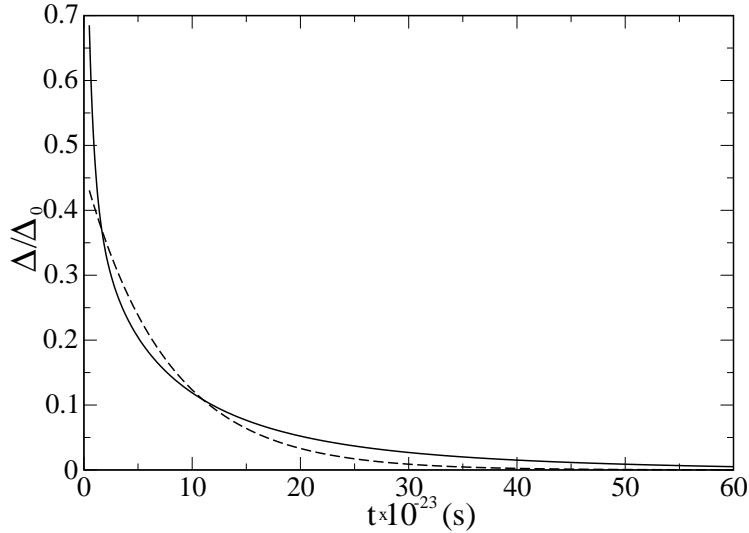}
\end{center}
\caption{Dependence of the mean square deviation $\Delta (t)$ from Eq. 
(\protect\ref{deltat}) on the time (solid line). The dashed line is the mean
square fit to the exponential function $\Delta (t)\sim \exp (-t/\protect\tau_{r})$.}
\label{fig5}
\end{figure}
Both curves in \textrm{Fig. 5} are normalized to the initial mean square
deviation
\[
\Delta _{0}=\int d\mathbf{p\ [}\delta f_{\mathrm{in}}(p)\mathbf{]}^{2},
\]%
where $\delta f_{\mathrm{in}}(p)=f_{\mathrm{in}}(p)-f_{\mathrm{eq}}(p)$. The
corresponding relaxation time $\tau _{r}$ is estimated as $\tau _{r}\approx
8.3\cdot 10^{-23}\mathrm{s}$. The obtained value of $\tau _{r}$ agrees with
a typical collisional relaxation time\ $\tau _{r,\mathrm{coll}}$ in a
nucleus $\tau _{r,\mathrm{coll}}=10^{-23}\div 10^{-22}\mathrm{s}$ \cite%
{bert78,shko05,khol82,wegm74}.

We note also that considering the time evolution of the distribution
function to the equilibrium (solid line in \textrm{Fig. 4}), we reach the
regime where the further change of distribution function is negligible. That
means that the collision integral in equation (\ref{4}) should be zero in
the case of equilibrium. To verify this fact we consider the time evolution
of the collision integral (\ref{stf6}). In \textrm{Fig. 6} we show the ratio%
\textbf{\ }%
\begin{equation}
R_{\mathrm{St}}(t)=\frac{\int d\mathbf{p\ [}\mathrm{St}\{f(t)\mathbf{\}]}^{2}%
}{\int d\mathbf{p\ [}\mathrm{St}\{f(t=0)\mathbf{\}]}^{2}},  \label{rst}
\end{equation}%
which represents the time dependence of the mean square of collision
integral $\mathrm{St}\{f(t)\}$\ normalized to the initial value of $\mathrm{%
St}\{f(t=0)\}$.

\begin{figure}[tbp]
\begin{center}
\includegraphics[scale=0.4,clip]{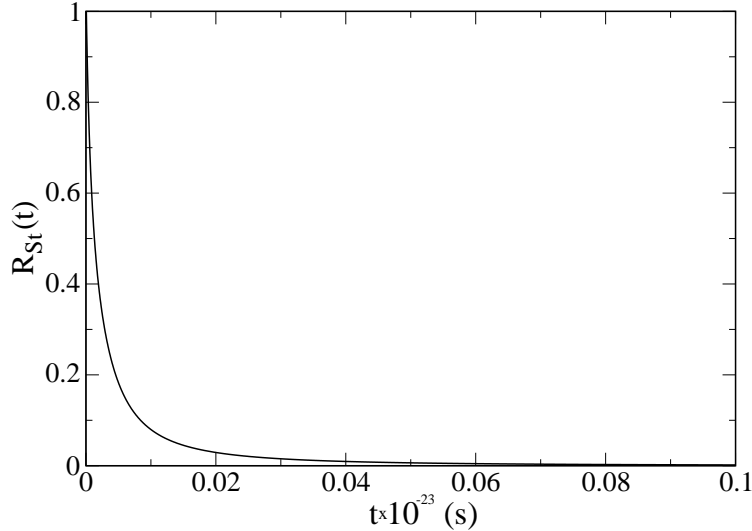}
\end{center}
\caption{Time dependence of the ratio $R_{St}(t)$.}
\end{figure}

As seen from \textrm{Fig. 6}, the collision term\textbf{\ }$\mathrm{St}%
\{f(t)\}$\textbf{\ }(\ref{stf6}) approaches to zero with time as it should
be in the equilibrium limit

\section{Conclusions}

We have considered the kinetic approach with the collision integral for the
Wigner distribution function using the diffusion approximation. We have
demonstrated that the collisional Landau-Vlasov kinetic equation can be
reduced to the form of the diffusion equation. We have established the
expressions for both the diffusion coefficient $D_{p}(p)$ and the drift term
$K_{p}(p)$ for the Fermi-systems where the relaxation processes occur on the
distorted Fermi surface.

We have found out that the simplest isotropic assumption for the collision
probability in the collision integral is insufficient for the correct
description of the kinetic coefficients. The forward scattering should be
ensured for the inter-particle interaction. For the description of the
inter-particle interaction we have used the finite-range potential. Such
kind of inter-particle interaction provides the necessary smallness of the
momentum transfer and thereby the forward scattering of particles.

For the nuclear Fermi-liquid we have calculated dependencies of the diffuse
coefficient $D_{p}(p)$ and the drift term $K_{p}(p)$ on the momentum $p$ and
the temperature $T$. For the diffuse coefficient $D_{p}(p)$ the momentum
dependence possesses the clearly observed maximum of order of $10^{-22}\
\mathrm{MeV}^{2}\mathrm{\cdot fm}^{-2}\mathrm{\cdot s}$ near the Fermi
momentum for different temperatures. With an increase of temperature the
value of the diffusion coefficient increases also. Excluding the small
momentum values, the drift term $K_{p}(p)$ has negative sign and the minimum
of order of $10^{-23}\ \mathrm{MeV\cdot fm}^{-2}\mathrm{\cdot s}$. Inside
the Fermi sphere for $p<p_{F}$ the drift term $K_{p}(p)$ shows the strong
dependence on the temperature. For $p>p_{F}$ the temperature dependence of $%
K_{p}(p)$ is practically negligible.

We have established the finite value of the diffusion coefficient in a cold
Fermi system with $D(p_{F},T=0)=2.1\cdot 10^{-22}$ $\mathrm{MeV}^{2}\cdot
\mathrm{fm}^{-2}\cdot \mathrm{s}$ which is caused by the relaxation on the
distorted Fermi-surface at $T=0$. The diffusion process was investigated
numerically assuming the initial non-equilibrium $1p1h$-excitation for a
finite Fermi-system with number of particles $A=16$. It was shown that the $%
1p1h$-excitation relaxes to the equilibrium Fermi distribution. The
numerical estimate for the relaxation time $\tau _{r}$ gives $\tau
_{r}\approx 8.3\cdot 10^{-23}s$ which is close to the corresponding
estimates in a nuclear Fermi-liquid obtained within the kinetic theory.

\appendix

\section{Moment representation for the collision integral.}

Applying the expansions of $W(\mathbf{p}_{1},\mathbf{p}_{3})$ and $%
\widetilde{W}(\mathbf{p}_{1},\mathbf{p}_{3})$ over small $\mathbf{s}=\mathbf{%
p}_{3}-\mathbf{p}_{1}$ (see Eqs. (\ref{w13}) and (\ref{w31})), the collision
integral (\ref{st1}) can be written as
\begin{eqnarray}
\mathrm{St}\{f \}&\approx &\int \frac{gd\mathbf{s}}{(2\pi \hbar )^{3}} \left[
\left( W(\mathbf{p}_{1},\mathbf{s})-\widetilde{W}(\mathbf{p}_{1}, \mathbf{s}%
)\right) \tilde{f}(\mathbf{p}_{1})f(\mathbf{p}_{1})\right.  \nonumber \\
&+& s_{\nu} \left\{ \left( W(\mathbf{p}_{1},\mathbf{s}) \tilde{f}(\mathbf{p}%
_{1}) +\widetilde{W}(\mathbf{p}_{1},\mathbf{s})f(\mathbf{p}%
_{1})\right)\nabla_{p_{1},\nu } f(\mathbf{p}_{1}) \right.  \nonumber \\
&+& \left. \frac{1}{2}\tilde{f}(\mathbf{p}_{1})f(\mathbf{p}%
_{1})\nabla_{p_{1},\nu } \left( W(\mathbf{p}_{1},\mathbf{s})- \widetilde{W}(%
\mathbf{p}_{1},\mathbf{s})\right) \right\}  \nonumber \\
&+& \frac{1}{2}s_{\nu }s_{\mu }\left\{ W(\mathbf{p}_{1},\mathbf{s}) \tilde{f}%
(\mathbf{p}_{1})\nabla _{p_{1},\nu }\nabla _{p_{1},\mu }f(\mathbf{p}%
_{1})\right.  \nonumber \\
&+&\left. \left( \tilde{f}(\mathbf{p}_{1})\nabla _{p_{1},\nu }W(\mathbf{p}%
_{1},\mathbf{s})+f(\mathbf{p}_{1})\nabla _{p_{1},\nu }\widetilde{W}(\mathbf{p%
}_{1},\mathbf{s})\right) \nabla _{p_{1},\mu }f(\mathbf{p}_{1})\right.
\nonumber \\
&+ & f(\mathbf{p}_{1}) \left( \widetilde{W}(\mathbf{p}_{1},\mathbf{s})
\nabla_{p_{1},\nu } \nabla_{p_{1},\mu } f(\mathbf{p}_{1}) \right.  \nonumber
\\
&+ & \left. \left. \left. \frac{1}{4}\tilde{f}(\mathbf{p}_{1})\nabla
_{p_{1},\nu }\nabla _{p_{1},\mu }\left( W(\mathbf{p}_{1},\mathbf{s})-%
\widetilde{W}(\mathbf{p}_{1},\mathbf{s})\right) \right) \right\} \right].
\label{stf1}
\end{eqnarray}
Using then Eqs. (\ref{gains}), \ (\ref{losses}) and (\ref{wdsdo}), we obtain
\begin{eqnarray}
W(\mathbf{P},\mathbf{s})&\simeq &\frac{2g}{m^{2}}\frac{d\sigma }{d\Omega } (%
\mathbf{s}^{2}) \int d\mathbf{p}_{2}d\mathbf{p}_{4}\tilde{f}(\mathbf{p}_{2})
f(\mathbf{p}_{4})\delta \left( \mathbf{p}_{2}-\mathbf{p}_{4}-\mathbf{s}%
\right)  \nonumber \\
&\times & \delta \left( \varepsilon _{2}-\varepsilon _{4} -\frac{\mathbf{%
P\cdot }\mathbf{s}}{m}\right)  \label{gain1}
\end{eqnarray}
and
\begin{eqnarray}
\widetilde{W}(\mathbf{P},\mathbf{s})&\simeq &\frac{2g}{m^{2}}\frac{d\sigma
} {d\Omega }(\mathbf{s}^{2}) \int d\mathbf{p}_{2}d\mathbf{p}_{4}\ f(\mathbf{p%
}_{2})\tilde{f}(\mathbf{p}_{4})\delta \left( \mathbf{p}_{2}-\mathbf{p}_{4}-%
\mathbf{s}\right)  \nonumber \\
&\times & \delta \left( \varepsilon _{2}-\varepsilon _{4} -\frac{\mathbf{%
P\cdot }\mathbf{s}}{m}\right).  \label{loss1}
\end{eqnarray}
The collision integral of Eq. (\ref{stf1}) is transformed to the following
form
\begin{eqnarray}
\mathrm{St}\{f\} &=&\int \frac{gd\mathbf{s}}{(2\pi \hbar )^{3}}\
s_{\nu}\left\{ W(\mathbf{p}_{1},\mathbf{s})\left( \tilde{f}(\mathbf{p}_{1})
-f(\mathbf{p}_{1})\right) \nabla_{p_{1},\nu }f(\mathbf{p}_{1}) + \tilde{f}(%
\mathbf{p}_{1})f(\mathbf{p}_{1}) \right.  \nonumber \\
&\times & \left. \nabla_{p_{1},\nu} W(\mathbf{p}_{1}, \mathbf{s})\right\} +
\frac{1}{2}\int \frac{gd\mathbf{s}}{(2\pi \hbar )^{3}}\ s_{\nu }s_{\mu}
\left\{ W(\mathbf{p}_{1},\mathbf{s})\tilde{f}(\mathbf{p}_{1})\nabla_{p_{1},%
\nu } \nabla_{p_{1},\mu }f(\mathbf{p}_{1}) \right.  \nonumber \\
&+& \nabla_{p_{1},\nu} \left( W(\mathbf{p}_{1},\mathbf{s})\right)
\nabla_{p_{1},\mu } f(\mathbf{p}_{1}) +\left. W(\mathbf{p}_{1},\mathbf{s})f(%
\mathbf{p}_{1})\nabla_{p_{1}, \nu} \nabla _{p_{1},\mu} f(\mathbf{p}%
_{1})\right\}.  \label{stf2}
\end{eqnarray}

Introducing the two first moments $A_{\nu }(\mathbf{p}_{1})$ and $B_{\nu \mu
}(\mathbf{p}_{1})$ of scattering probability $W(\mathbf{p}_{1},\mathbf{s})$
as
\begin{equation}
A_{\nu }(\mathbf{p}_{1})=\int \frac{gd\mathbf{s}}{(2\pi \hbar )^{3}}\ s_{\nu
}\ W(\mathbf{p}_{1},\mathbf{s}),\quad B_{\nu \mu }(\mathbf{p}_{1})=\frac{1}{2%
}\int \frac{gd\mathbf{s}}{(2\pi \hbar )^{3}}\ s_{\nu }s_{\mu }\ W(\mathbf{p}%
_{1},\mathbf{s}),  \label{an}
\end{equation}%
the collision integral (\ref{stf2}) is reduced as%
\begin{equation}
\mathrm{St}\{f\}=\nabla _{p_{1},\nu }\left[ A_{\nu }(\mathbf{p}_{1})f(%
\mathbf{p}_{1})\tilde{f}(\mathbf{p}_{1})+B_{\nu \mu }(\mathbf{p}_{1})\nabla
_{p_{1},\mu }f(\mathbf{p}_{1})\right] .  \label{stf4}
\end{equation}%
Using the relations
\begin{eqnarray}
&&B_{\nu \mu }(\mathbf{p}_{1})\nabla _{p_{1},\mu }f(\mathbf{p}_{1})
\nonumber \\
&=&\nabla _{p_{1},\mu }\left( B_{\nu \mu }(\mathbf{p}_{1})\tilde{f}(\mathbf{p%
}_{1})f(\mathbf{p}_{1})\right) -\tilde{f}(\mathbf{p}_{1})f(\mathbf{p}%
_{1})\nabla _{p_{1},\mu }B_{\nu \mu }(\mathbf{p}_{1})+B_{\nu \mu }(\mathbf{p}%
_{1})\nabla _{p_{1},\mu }f^{2}(\mathbf{p}_{1})  \nonumber \\
&=&\nabla _{p_{1},\mu }\left( B_{\nu \mu }(\mathbf{p}_{1})f(\mathbf{p}%
_{1})\right) -f^{2}(\mathbf{p}_{1})\nabla _{p_{1},\mu }B_{\nu \mu }(\mathbf{p%
}_{1})-\tilde{f}(\mathbf{p}_{1})f(\mathbf{p}_{1})\nabla _{p_{1},\mu }B_{\nu
\mu }(\mathbf{p}_{1}),  \label{rel1}
\end{eqnarray}%
we rewrite the collision integral (\ref{stf4}) as
\begin{eqnarray}
\mathrm{St}\{f\} &=&-\ \nabla _{p_{1},\nu }\left[ \left\{ \nabla_{p_{1},\mu}
B_{\nu\mu}(\mathbf{p}_{1})-A_{\nu }(\mathbf{p}_{1})\right\} f(\mathbf{p}_{1})%
\tilde{f}(\mathbf{p}_{1})+f^{2}(\mathbf{p}_{1})\nabla_{p_{1},\mu} B_{\nu\mu}(%
\mathbf{p}_{1})\right]  \nonumber \\
&+&\nabla _{p_{1},\nu }\nabla _{p_{1},\mu }(f(\mathbf{p}_{1})B_{\nu\mu} (%
\mathbf{p}_{1})).  \label{stf5}
\end{eqnarray}
Finally, we obtain
\begin{eqnarray}
\mathrm{St}\{f\} &=&-\ \nabla _{p_{1},\nu }\left[ K_{p}(\mathbf{p}_{1})
(\nabla_{p_{1},\nu}\varepsilon_{p_{1}})f(\mathbf{p}_{1}) \tilde{f}(\mathbf{p}%
_{1})+f^{2}(\mathbf{p}_{1})\nabla _{p_{1},\nu} D_{p}(\mathbf{p}_{1})\right]
\nonumber \\
&+&\nabla_{p_{1},\nu}^{2}\,\left[ f(\mathbf{p}_{1})\,D_{p}(\mathbf{p}_{1})%
\right].  \label{stf6}
\end{eqnarray}

\section{Diffuse and drift terms}

Integrating in Eq. (\ref{dp}) over $\mathbf{s}$, one obtains
\begin{eqnarray}
D_{p}(\mathbf{p}_{1}) &\approx &\frac{g^{2}r_{0}^{6}v_{0}^{2}}{48\pi
^{2}\hbar ^{7}}\int d\mathbf{p}_{2}d\mathbf{p}_{4}\ (\mathbf{p}_{2}-\mathbf{p%
}_{4})^{2} \exp \left( -4(\mathbf{p}_{2}-\mathbf{p}_{4})^{2}r_{0}^{2}/\hbar
^{2}\right) \tilde{f}(\mathbf{p}_{2})f(\mathbf{p}_{4})  \nonumber \\
&\times &\delta \left( \varepsilon _{2}-\varepsilon _{4}-\frac{\mathbf{p}_{1}%
}{m} (\mathbf{p}_{2}-\mathbf{p}_{4})\right).  \label{dp1}
\end{eqnarray}
The argument of the delta function in Eq. (\ref{dp1}) can be written as
\[
\varepsilon _{2}-\varepsilon _{4}-\frac{\mathbf{p}_{1}}{m}(\mathbf{p}_{2}-
\mathbf{p}_{4})=\frac{1}{2m}\left( (\mathbf{p}_{2}-\mathbf{p}_{1})^{2} -(%
\mathbf{p}_{4}-\mathbf{p}_{1})^{2}\right) ,
\]
where $\varepsilon_{j}=\mathbf{p}_{j}^{2}/2m$. We will introduce the new
variables $\mathbf{p}_{2}-\mathbf{p}_{1}=\mathbf{q}$ and $\mathbf{p}_{4}-%
\mathbf{p}_{1}=\mathbf{k}$ which allows one to rewrite Eq. (\ref{dp1}) in
the following form
\begin{eqnarray}
D_{p}(\mathbf{p}_{1})&\approx &\frac{g^{2}mr_{0}^{6}v_{0}^{2}}{24\pi ^{2}
\hbar^{7}} \int d\mathbf{q}\ d\mathbf{k}\ (\mathbf{q}-\mathbf{k})^{2}\exp
\left( -4(\mathbf{q}-\mathbf{k})^{2}r_{0}^{2}/\hbar ^{2}\right) \tilde{f}(%
\mathbf{q} +\mathbf{p}_{1})f(\mathbf{k}+\mathbf{p}_{1})  \nonumber \\
&\times & \delta \left( \mathbf{q}^{2}-\mathbf{k}^{2}\right) .  \label{dp3}
\end{eqnarray}
Using the spherically symmetric distribution function $f(\mathbf{p})$ and
the relation
\[
\delta \left( x^{2}-a^{2}\right) =\frac{\delta (x-a)+\delta (x+a)}{2|a|},
\]%
we will rewrite Eq. (\ref{dp3}) as
\begin{eqnarray}
D_{p}(\mathbf{p}_{1}) &\approx &\frac{g^{2}mr_{0}^{6}v_{0}^{2}}{48\pi
^{2}\hbar ^{7}}\int_{0}^{\infty }q^{2}dq\ \int d\Omega _{q}\int_{0}^{\infty
}kdk\ \int d\Omega _{k}\left( q^{2}+k^{2}-2qk\cos \widehat{\mathbf{q}\mathbf{%
k}}\right)  \nonumber \\
&\times &\exp \left( -4(q^{2}+k^{2}-2qk\cos \widehat{\mathbf{q}\mathbf{k}}%
)r_{0}^{2}/\hbar ^{2}\right) \left\{ \delta (q-k)+\delta (q+k)\right\}
\nonumber \\
&\times &\tilde{f}\left( \sqrt{q^{2}+p_{1}^{2}+2qp_{1}\cos \widehat{\mathbf{q%
}\mathbf{p}_{1}}}\right) f\left( \sqrt{k^{2}+p_{1}^{2}+2kp_{1}\cos \widehat{%
\mathbf{k}\mathbf{p}_{1}}}\right) .  \label{dp4y}
\end{eqnarray}%
Integrating in Eq. (\ref{dp4y}) over $q$, we obtain
\begin{eqnarray}
D_{p}(\mathbf{p}_{1}) &\approx &\frac{g^{2}mr_{0}^{6}v_{0}^{2}}{%
24\pi^{2}\hbar^{7}} \int_{0}^{\infty }dk\ k^{5}\int d\Omega _{k}\int
d\Omega_{q}\ \left( 1-\cos \widehat{\mathbf{q}\mathbf{k}}\right)  \nonumber
\\
&\times & \exp \left(-8k^{2}r_{0}^{2}(1-\cos \widehat{\mathbf{q}\mathbf{k}}%
)/\hbar ^{2}\right) \tilde{f}\left( \sqrt{k^{2}+p_{1}^{2}+2kp_{1}\cos
\widehat{\mathbf{q\cdot }\mathbf{p}_{1}}}\right)  \nonumber \\
&\times & f\left( \sqrt{k^{2}+p_{1}^{2}+2kp_{1}\cos \widehat{\mathbf{k\cdot }
\mathbf{p}_{1}}}\right).  \label{dp6x}
\end{eqnarray}

Using the addition theorem for spherical harmonics, we can write in an
arbitrary spherical coordinate system
\begin{eqnarray}
\cos \widehat{\mathbf{q\cdot }\mathbf{k}} &=&\cos \theta _{q}\cos \theta
_{k}+\sin \theta _{q}\sin \theta _{k}\cos (\phi _{q}+\phi _{k}),  \nonumber
\\
\cos \widehat{\mathbf{q\cdot }\mathbf{p}_{1}} &=&\cos \theta _{q}\cos \theta
_{p_{1}}+\sin \theta _{q}\sin \theta _{p_{1}}\cos (\phi _{q}+\phi _{p_{1}}),
\nonumber \\
\cos \widehat{\mathbf{k\cdot }\mathbf{p}_{1}} &=&\cos \theta _{k}\cos \theta
_{p_{1}}+\sin \theta _{k}\sin \theta _{p_{1}}\cos (\phi _{k}+\phi _{p_{1}}).
\nonumber
\end{eqnarray}%
Taking into account the spherically symmetry of the distribution functions $%
f(\mathbf{p}_{j})=f(p_{j})$ and using
\[
\cos \widehat{\mathbf{q\cdot }\mathbf{p}_{1}}=\cos \theta _{q},\qquad \cos
\widehat{\mathbf{k\cdot }\mathbf{p}_{1}}=\cos \theta _{k},
\]%
we obtain
\begin{eqnarray}
D_{p}(p_{1}) &\approx &\frac{g^{2}mr_{0}^{6}v_{0}^{2}}{24\pi ^{2}\hbar ^{7}}%
\int_{0}^{\infty }dk\ k^{5}\int d\Omega _{k}\int d\Omega _{q}\ \left( 1-\cos
\widehat{\mathbf{q}\mathbf{k}}\right)  \nonumber \\
&\times & \exp \left( -8k^{2}r_{0}^{2}(1-\cos \widehat{\mathbf{q}\mathbf{k}}%
)/\hbar ^{2}\right) \tilde{f}\left( \sqrt{k^{2}+p_{1}^{2}+2kp_{1}\cos \theta
_{q}}\right)  \nonumber \\
&\times & f\left( \sqrt{k^{2}+p_{1}^{2}+2kp_{1}\cos \theta _{k}}\right) .
\label{dp7x}
\end{eqnarray}%
We will also introduce the following notations
\[
j(\alpha )=\int_{0}^{2\pi }d\phi _{q}\int_{0}^{2\pi }d\phi _{k}\ \exp \left(
\alpha (\phi _{q}+\phi _{k})\right) ,
\]
\[
\widetilde{j}(\alpha )=\int_{0}^{2\pi }d\phi _{q}\int_{0}^{2\pi }d\phi _{k}\
\cos (\phi _{q}+\phi _{k})\exp \left( \alpha (\phi _{q}+\phi _{k})\right) ,
\]%
which are attributed as functions of the argument $\alpha $. Using $%
j(\alpha) $ and $\widetilde{j}(\alpha )$, one can write the angular
integrals in Eq. (\ref{dp7x}) as
\begin{eqnarray*}
&& \int_{0}^{2\pi }d\phi _{q}\int_{0}^{2\pi }d\phi _{k}\ \left( 1-\cos
\widehat{\mathbf{q}\mathbf{k}}\right) \exp \left( -8k^{2}r_{0}^{2}(1-\cos
\widehat{\mathbf{q}\mathbf{k}})/\hbar ^{2}\right) \\
&=&\exp \left(-8k^{2}r_{0}^{2}(1-\cos \theta _{q}\cos \theta _{k})/\hbar
^{2}\right) \\
&\times & \left\{ (1-\cos \theta _{q}\cos \theta _{k})j(8k^{2}r_{0}^{2}\sin
\theta _{q}\sin \theta _{k}/\hbar ^{2})-\sin \theta _{q}\sin \theta _{k}\
\widetilde{j}(8k^{2}r_{0}^{2}\sin \theta _{q}\sin \theta
_{k}/\hbar^{2})\right\}
\end{eqnarray*}
Finally, we obtain the diffuse coefficient $D_{p}(p)$ in the form given by
Eq. (\ref{dp8x}).

The drift coefficient $K_{p}(\mathbf{p})$ of Eq. (\ref{kpdef1}) can be
reduced similarly to above described procedure. Using derivation of Eq. (\ref%
{A1}), we will reduce $A(\mathbf{p}_{1})$ to the following form
\begin{eqnarray*}
A(\mathbf{p}_{1})&\approx &\frac{g^{2}mr_{0}^{6}v_{0}^{2}}{4\pi
^{2}\hbar^{7}p_{1}} \int d\mathbf{q}\ d\mathbf{k}\ \mathbf{p}_{1}(\mathbf{q}-%
\mathbf{k}) \exp \left( -4(\mathbf{q}-\mathbf{k})^{2}r_{0}^{2}/\hbar
^{2}\right) \tilde{f}(\mathbf{q}+\mathbf{p}_{1})f(\mathbf{k}+\mathbf{p}_{1})
\\
&\times & \delta \left( \mathbf{q}^{2}-\mathbf{k}^{2}\right),
\end{eqnarray*}
or
\begin{eqnarray}
A(p_{1}) &\approx &\frac{g^{2}mr_{0}^{6}v_{0}^{2}}{8\pi ^{2}\hbar ^{7}}
\int_{0}^{\infty }dk\ k^{4}\int d\Omega _{k}\int d\Omega _{q}\ \left( \cos
\theta _{q}-\cos \theta _{k}\right)  \nonumber \\
&\times & \exp \left( -8k^{2}r_{0}^{2}(1-\cos \widehat{\mathbf{q}\mathbf{k}}%
)/\hbar^{2}\right) \tilde{f}\left( \sqrt{k^{2}+p_{1}^{2}+2kp_{1}\cos \theta
_{q}}\right)  \nonumber \\
&\times & f\left( \sqrt{k^{2}+p_{1}^{2}+2kp_{1}\cos \theta_{k}}\right).
\label{dk7x}
\end{eqnarray}%
Finally, we obtain
\begin{eqnarray}
A(p) &\approx &\frac{g^{2}mr_{0}^{6}v_{0}^{2}}{8\pi ^{2}\hbar ^{7}}%
\int_{0}^{\infty }dk\ k^{4}\int_{-1}^{1}dx\int_{-1}^{1}dy\ (x-y)  \nonumber
\\
&\times &\exp \left( -8k^{2}r_{0}^{2}(1-xy)/\hbar ^{2}\right) j\left(
8k^{2}r_{0}^{2}\sqrt{1-x^{2}}\sqrt{1-y^{2}}/\hbar ^{2}\right)  \nonumber \\
&\times &\tilde{f}\left( \sqrt{k^{2}+p^{2}+2kp_{1}x}\right) f\left( \sqrt{%
k^{2}+p^{2}+2kpy}\right) .  \label{dk8x}
\end{eqnarray}


\begin{thebibliography}{99}
\bibitem{kosh04} V.M. Kolomietz and S. Shlomo, Phys. Rep. \textbf{390}, 133
(2004).

\bibitem{bert78} G. Bertsch, Z. Phys. \textbf{A289}, 103 (1978).

\bibitem{kopl95} V.M. Kolomietz, V.A. Plujko and S. Shlomo, Phys. Rev. C
\textbf{52}, 2480 (1995).

\bibitem{mako95} A.G. Magner, V.M. Kolomietz, H. Hofmann and S.Shlomo, Phys.
Rev. \textbf{C51}, 2457 (1995).

\bibitem{kiko96} D. Kiderlen, V.M. Kolomietz and S. Shlomo, Nucl. Phys.
\textbf{A608}, 32 (1996).

\bibitem{kopl96} V.M. Kolomietz, V.A. Plujko and S.Shlomo, Phys. Rev. C
\textbf{54}, 3014 (1996).

\bibitem{diko99} M. Di Toro, V.M. Kolomietz and A.B. Larionov, Phys. Rev.
\textbf{C59}, 3099 (1999).

\bibitem{lipi81} E.M. Lifshitz and L.P. Pitaevskii, \textit{Physical Kinetics}, 
Pergamon Press, Oxford, 1981, Ch. 2.

\bibitem{kaba62} L.P. Kadanoff and G. Baym, \textit{Quantum Statistical
Mechanics}, Benjamin, London, 1962, Ch. 9.

\bibitem{risc80} P. Ring and P. Schuck, \textit{The Nuclear Many-Body Problem}, 
Springer-Verlag, New York, 1980, Ch. 13.

\bibitem{abkh59} A.A. Abrikosov and I.M. Khalatnikov, Rep. Prog. Phys.
\textbf{22}, 329 (1959).

\bibitem{shda03} L. Shi and P. Danielewicz, Phys.Rev. C \textbf{68}, 064604
(2003).

\bibitem{coly11} D.D.S. Coupland, W.G. Lynch, M.B. Tsang, P. Danielewicz and
Yingxun Zhang, Phys. Rev. C \textbf{84}, 054603 (2011).

\bibitem{Dav.b.65} A.S. Davydov, \textit{Quantum Mechanics}, Pergamon Press,
Oxford, 1965.

\bibitem{wols82} G. Wolshin, Phys. Rev. Lett. \textbf{48}, 1004 (1982).

\bibitem{bomo1} A. Bohr and B.R. Mottelson, \textit{Nuclear Structure}, W.
A. Benjamin, New York, 1969, Vol. 1.

\bibitem{shko05} S. Shlomo and V.M. Kolomietz, Rep. Prog. Phys. \textbf{68},
1 (2005).

\bibitem{khol82} H. S. K\"{o}hler, Nucl. Phys. \textbf{A378}, 159 (1982).

\bibitem{wegm74} G. Wegmann, Phys. Lett. \textbf{B50}, 327 (1974).
\end{thebibliography}
\end{document}